\documentclass[letterpaper]{article} 
\usepackage{aaai23}  
\usepackage{times}  
\usepackage{bbding}
\usepackage{graphicx}
\usepackage{amsfonts}
\usepackage{subfigure}
\usepackage{amsmath}
\usepackage{multirow}
\usepackage{helvet}  
\usepackage{courier}  
\usepackage[hyphens]{url}  
\usepackage{graphicx} 
\usepackage{natbib}  
\usepackage{caption} 
\usepackage{algorithm}
\usepackage{algorithmic}
\usepackage{newfloat}
\usepackage{listings}
\DeclareCaptionStyle{ruled}{labelfont=normalfont,labelsep=colon,strut=off} 
\floatstyle{ruled}
\newfloat{listing}{tb}{lst}{}
\floatname{listing}{Listing}
\title{Selector-Enhancer: Learning Dynamic Selection of Local and Non-local Attention Operation for Speech Enhancement}
\author{
    Xinmeng Xu$^{1}$, Weiping Tu$^{1,2,3}$\thanks{This work was supported in part by the National Nature Science Foundation of China (No. 62071342, No.62171326), the Special Fund of Hubei Luojia Laboratory (No. 220100019), the Hubei Province Technological Innovation Major Project (No. 2021BAA034) and the Fundamental Research Funds for the Central Universities (No. 2042022kf0001). \textit{(Corresponding Author: Weiping Tu)}.}, Yuhong Yang$^{1,3}$
}
\affiliations{
    $^1$National Engineering Research Center for Multimedia Software, School of Computer Science, Wuhan University,  China\\
    $^2$Hubei Luojia Laboratory, China \\
    $^3$Hubei Key Laboratory of Multimedia and Network Communication Engineering, Wuhan University, China\\
    $\{$xuxinmeng, tuweiping, yangyuhong$\}$whu.edu.cn
}

\begin{document}

\maketitle

\begin{abstract}
Attention mechanisms, such as local and non-local attention, play a fundamental role in recent deep learning based speech enhancement (SE) systems. However, natural speech contains many fast-changing and relatively brief acoustic events, therefore, capturing the most informative speech features by indiscriminately using local and non-local attention is challenged. We observe that the noise type and speech feature vary within a sequence of speech and the local and non-local operations can respectively extract different features from corrupted speech. To leverage this, we propose Selector-Enhancer, a dual-attention based convolution neural network (CNN) with a feature-filter that can dynamically select regions from low-resolution speech features and feed them to local or non-local attention operations. In particular, the proposed feature-filter is trained by using reinforcement learning (RL) with a developed difficulty-regulated reward that is related to network performance, model complexity, and “the difficulty of the SE task”. The results show that our method achieves comparable or superior performance to existing approaches. In particular, Selector-Enhancer is potentially effective for real-world denoising, where the number and types of noise are varies on a single noisy mixture.
\end{abstract}

\section{Introduction}
Speech enhancement (SE) aims at separating the underlying high quality and intelligently speech $\textbf{s}$ from a noise-corrupted speech signal $\textbf{y} = \textbf{s} + \textbf{n}$, where $\textbf{n}$ represents additive noise. For many applications such as mobile communication and hearing aids, SE algorithms are restricted to operate with low latency and often also to use only single-channel inputs. With the recent advances in supervised learning, deep neural networks (DNNs) have become state-of-the-art for single-channel SE. They typically operate in the short-time Fourier transform (STFT) domain and estimate the target clean speech from the noisy signal via direct spectral mapping \cite{ref1} or time-frequency masking \cite{ref2}.

Several studies have shown the efficiency of local processing by convolutional neural networks (CNNs) for STFT-domain SE, which are structurally well-suited to focus on local patterns such as harmonic structures in the speech spectra \cite{ref3}. However, the speech also exhibits non-linguistic long-term dependencies, such as gender, dialect, speaking rating, and emotional state \cite{ref4}. For aggregating more informative speech features, several strategies are proposed, such as encoder-decoder architecture \cite{ref3}, dilated convolutions \cite{ref7} or applying long-short term memory (LSTM) networks \cite{ref28, ref17}, to enlarge the receptive field. 

\begin{figure}[t]
  \centering
  \includegraphics[width=0.98\linewidth]{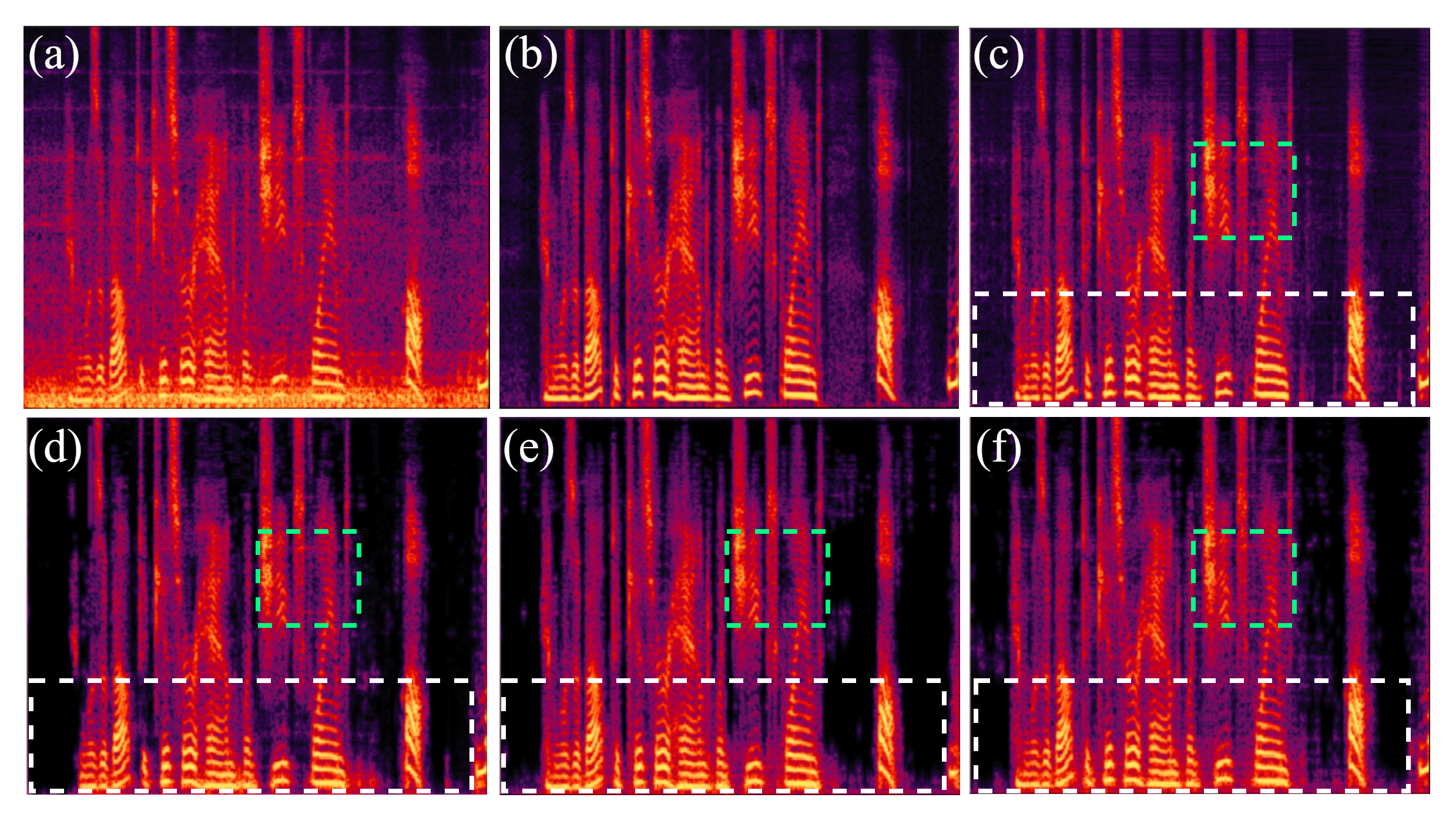}
  \caption{Comparison with (c) local and non-local attention with feature-filter, (d) local attention, (e) non-local attention, and (f) local and non-local attention based SE model enhanced speech spectrogram, in which most obvious details of spectrogram distinction are framed by a gray and green dotted box. Note that (a) is a noisy speech spectrogram, and (b) is a clean speech spectrogram.}
  \label{fig:0}
\end{figure}

Furthermore, the non-local operation \cite{ref9}, which takes the advantage of self-similarity within the whole speech to recover a local speech segment, is effective to establish the long-range dependence based on time-frequency (T-F) units of speech spectra. However, T-F units of noisy speech are usually corrupted, therefore, establishing the correlation between T-F units is unreliable, especially when speech is heavily corrupted. Besides, the non-local operation has $\mathcal{O}(T^2)$ computational complexity for a sequence length of $T$, which increases the computational burden for the SE model, hence the practical usage is limited. In addition to non-local attention, local attention processes the local neighborhood, which scales more informative feature regions with ``high attention'' while putting ``low attention'' to other feature regions and achieving a linear complexity to the input feature. However, local attention is unable to extract speech global information that is important to speech enhancement tasks, due to the limited receptive field.

According to the above analysis, local and non-local attentions have shortcomings and merits. Figure~\ref{fig:0}(d) and Figure~\ref{fig:0}(e) show an example enhanced spectrogram processed by local and non-local attention-based SE models. The local attention-based speech enhancement model failed to process the noise component in the green dotted box. In contrast, the non-local attention-based system insufficiently removes the noise component in the white dotted box. However, integrating both attentions indiscriminately is a challenge to keep their merits while avoiding their shortcomings. For example, as shown in Figure ~\ref{fig:0}(f), the SE system by directly applying local and non-local attention operations still failed to remove the noise component in the green box while remaining the low-frequency noise component in the white box.

In this study, we propose Selector-Enhancer, a novel framework that jointly processes different regions of a noisy speech spectrogram via local and non-local attention operation with the guidance of adjacent features. Particularly, a feature-filter is proposed and is trained together with dual-attention based CNN. The feature-filter first predicts a low-resolution dispatch policy mask, in which each time-frequency (T-F) unit represents a policy for a region in the input noisy spectrogram. After that, the features of different regions are sent to specific attention operations (local or non-local attention) according to the corresponding policy on the predicted mask. Since the feature selection is non-differentiable, the feature filter is trained in a reinforcement learning (RL) framework driven by a reward to achieve a good complexity-performance trade-off. 

In a nutshell, the contributions of our work are three-fold\footnote{Code is available at: https://github.com/XinmengXu/Selector\\-Enhancer.}
\begin{itemize}
\item We propose a novel SE framework for making a trade-off between local and non-local attention operations by introducing a feature-filter that predicts a policy mask to select the feature of different spectrogram regions and to send to specific attention operations.

\item We develop a difficulty-regulated reward, which is related to performance, complexity, and “the difficulty of the SE task”. Specifically, the difficulty is made by the loss function, since the high loss denotes the high difficulty of the SE task.

\item The proposed Selector-Enhancer achieves comparable and superior performance to recent approaches with faster speed and is particularly effective when noise distribution is variant within a speech sequence. 
\end{itemize}

\section{Related Work}
\subsection{Non-local Attention}
Long-range dependencies are important prior to SE, and this prior can be effectively obtained by non-local attention operation, which calculates the mutual similarity between T-F units in each frame, which is helpful for capturing the global information in the frequency domain with a slightly increased computing complexity. In general, non-local attention is defined as follows \cite{ref11}:
\begin{equation}
    y_i = \frac{1}{\mathcal{C}(x)}\sum_{\forall j}f(x_i, x_j)g(x_j),
\end{equation}
where $x$ and $y$, respectively, represent the input and output tensor of the operation with the same size, $f$ denotes the pairwise function to calculate the correlation between the locations of the feature map, $g$ signifies the unary input function for information transform, and $\mathcal{C}(x)$ is a normalization factor.

Recently, deep neural networks (DNNs) have been prevalent in the community of SE, and non-local attention-based SE models have been proposed for modeling long-term dependencies of speech signals. In \cite{ref10}, embedding function $g$ is a convolutional layer that can be viewed as a linear embedding function, and dot product, i.e., $\phi(\textbf{y}_i, \textbf{y}_j)=\textbf{y}_i^{\top}\textbf{y}_j$ is applied to compute the similarity between query and keys. However, non-local attention is unreliable when speech is heavily corrupted and colored noise is contained in the background mixture. In addition, a transformer with Gaussian-weighted non-local attention is proposed \cite{ref15}, whose attention weights are attenuated according to the distance between correlated T-F units. The attenuation is determined by the Gaussian variance which can be learned during training. While all these methods demonstrate the benefits of non-local attention, their approach is unreliable when speech is heavily corrupted and has high computational costs.

\begin{figure*}
  \centering
  \includegraphics[width=0.90\linewidth]{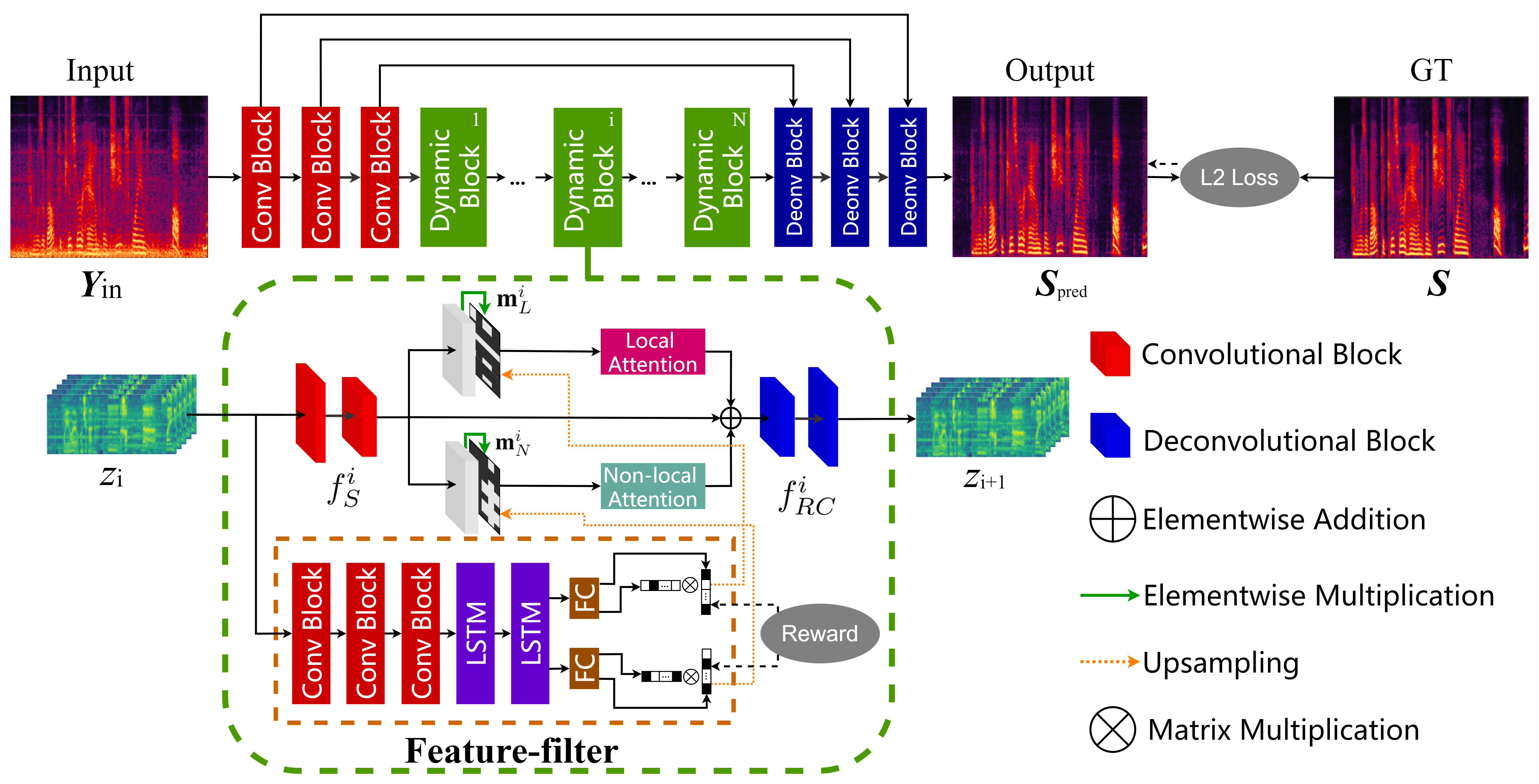}
  \caption{The proposed Selector-Enhancer, which is composed of a dual-attention based CNN and a feature filter. The dual-attention-based CNN contains $N$ dynamic blocks, each of which has 2 optional paths. The feature filter is depicted at the bottom left corner. It predicts a policy mask with two channels, both of which are then up-sampled by nearest-neighbor interpolation to select eligible regions for the corresponding paths.}
  \label{fig:1}
\end{figure*}

\subsection{Local Attention}
Local attention focuses on the important speech components in an audio stream with ``high attention'' while perceiving the unimportant region (e.g., noise or interference) in ``low attention'', according to the past and current time frame, thus adjusting the focal point over time. In \cite{ref16}, a lightweight casual transformer with local self-attention is proposed for real-time SE in computation resource-limited environments. Local attention based RNN-LSTM \cite{ref17} is proposed for superior SE performances by modeling important sequential information, in which the RNN model learns the weights of past input features implicitly when predicting an enhanced frame, the attention mechanism calculates correlations between past frames and the current frame to be enhanced and give weights to past frames explicitly. Compared with non-local attention, local attention is more suitable for real-time settings and more lightweight \cite{ref16, ref17}, since they do not rely on future information. However, compared with non-local operations, the main drawback of the local attention mechanism during SE is the relatively small receptive field.

\subsection{Dynamic Networks}
Dynamic networks are effective to achieve a trade-off between speech and performance in different tasks. Recently, several approaches \cite{ref20, ref43, ref44} are applied to dynamic networks to reduce the computational cost in ResNet by dynamically skipping some residual blocks. Although the aforementioned methods mainly focus on a single task, a routing network \cite{ref22} is developed to facilitate multi-task learning, and their method seeks an appropriate network path for each specific task. Existing dynamic networks successfully explore the route policy that depends on either the task to address. However, we aim at selecting different regions of input speech spectrogram to send to specific attention operations, in which noise information (noise type and signal-to-noise ratio (SNR)) and speech information (speaker identity and speech content) have a weighty influence on the performance and they should be both considered in the selection operation. 

\section{Methodology}
Our task is to recover a clean speech $\textbf{s}$ from a noisy observation $\textbf{y}\in \mathbb{R}^{T\times F \times 2}$. We propose Selector-Enhancer that can process each noisy observation according to its speech and noise features through a specific attention mechanism. Figure~\ref{fig:1} gives a detailed illustration of the Selector-Enhancer framework. The Selector-Enhancer takes the noisy raw waveform and firstly transforms into a short-time Fourier transform (STFT) spectrogram, denoted by $\textbf{Y}_{\text{in}}$ in Figure~\ref{fig:1}. The output of Selector-Enhancer is clean spectrogram $\textbf{S}_{\text{pred}}$, and then transforms into raw waveform through inverse STFT.

\subsection{Architecture of Selector-Enhancer}
We aim to design a framework that can offer different attention mechanism options for SE. To this end, as shown in Figure~\ref{fig:1}, Selector-Enhancer is composed of 3 convolutional blocks at the start- and end-point for extracting features and reconstructing the target speech, and $N$ dynamic blocks in the middle. Each convolutional block contains a stacked 2D convolutional layer which is followed by batch normalization, and exponential linear unit (ELU). Additionally, skip connections are utilized to concatenate the output of each convolutional block to the input of the corresponding deconvolutional block.

In the $i$-th dynamic block $f^i_{DB}$, there is a shared path $f_S^i$ containing two convolutional blocks for extracting deeper speech features, which every feature map should pass through. Paralleling the shared path, a feature-filter generates a probabilistic distribution of plausible paths for selection by each T-F unit of the input feature map. Following the shared path, there are two paths for local and non-local attention denoted by $f^i_{LA}$ and $f^i_{NA}$. According to the output of the feature-filter, the path with the highest probability is activated, where the path index is represented by action $\alpha_i$. Since the path selection is for every T-F unit rather than a feature map, the action $\alpha_i$ is a two-channel mask rather than a scalar. The first channel $m^i_{L}$ implies the regions that go through the path for local attention, and similarly, the second channel $m^i_{N}$ corresponds to the path for non-local attention. The whole process is formulated as:
\begin{equation}
\begin{aligned}
     z_{i+1} = f^i_R &(f^i_{LA}(f_S^i(z_i)\cdot m^i_{L}) \\
     &+ f^i_{NA}(f_S^i(z_i)\cdot m^i_{N}) + f_S^i(z_i)),
\end{aligned}
\end{equation}
where $z_i$ and $z_{i+1}$ represent the input and output of the $i$-th dynamic block, respectively, and $f^i_R$ denotes the decoder of dynamic block. Note that ``$\cdot$'' denotes the element-wise multiplication, and each channel of feature $f_S^i(z_i)$ performs the same element-wise multiplication with mask $m^i$. Specially, the mask $m^i$ utilizes two branches to generate a 1-D frequency-dimension mask $\textbf{F}_M \in \mathbf{R}^{1\times F}$ and a 1-D time-frame mask $\textbf{T}_M \in \mathbf{R}^{T\times 1}$ in parallel, then combines them with a matrix multiplication to obtain the final 2-D mask.

\begin{figure}
\centerline{\includegraphics[width=\linewidth]{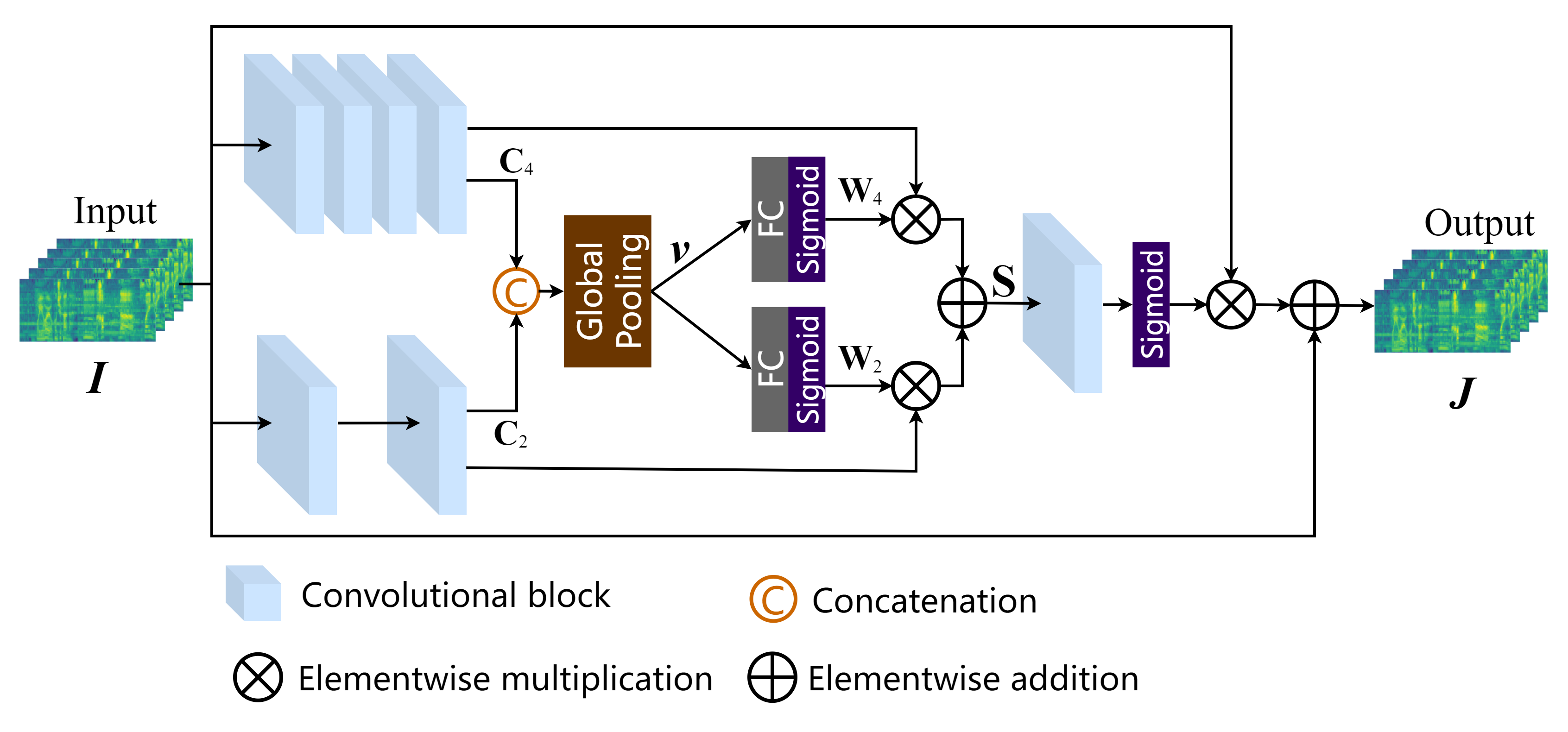}}
\caption{Block Diagram of Local Attention.}
\label{fig:2}
\end{figure}

\begin{figure}
\centerline{\includegraphics[width=0.9\linewidth]{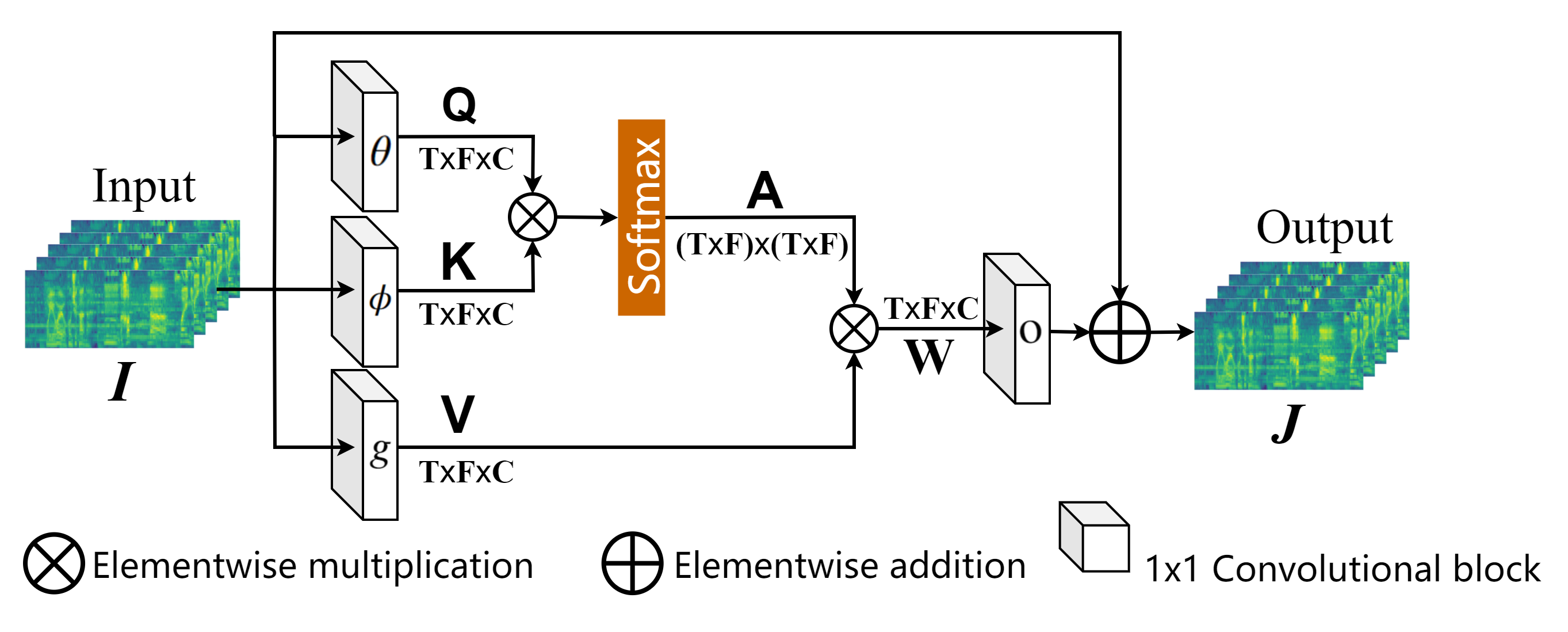}}
\caption{Block Diagram of Non-local Attention.}
\label{fig:3}
\end{figure}

\subsection{Local and Non-local Attention}
The local and non-local attention operations are demonstrated as follows:

\noindent \textbf{Local Attention.} \quad Inspired by \cite{ref23}, we employ a channel-wise attention mechanism to design the local attention to perform channel selection with multiple sizes of the receptive field. The detailed architecture of the local attention is shown in Figure~\ref{fig:2}. In the local attention, we utilized two branches to carry different numbers of convolution layers to generate feature maps with different sizes of receptive filed. The channel-wise attention is independently performed on these two outputs. As illustrated in Figure~\ref{fig:2}, given input feature $\textbf{I}$, two-layer convolutional blocks and four-layer convolutional blocks generate their output $\textbf{C}_2$ and $\textbf{C}_4$ in a paralleled manner:
\begin{equation}
    \left\{
    \begin{array}{lr}
         \textbf{C}_2 = f_2(\textbf{I}) \\
         \textbf{C}_4 = f_4(\textbf{I}),
    \end{array}
    \right.
\end{equation}
where $f_2(\cdot)$ and $f_4(\cdot)$ refer to functions of the two and four convolutional blocks. After getting $\textbf{C}_2$ and $\textbf{C}_4$, the scaled feature $S$ is obtained in the following steps:
\begin{equation}
    \left\{
    \begin{array}{lr}
         \textbf{v} = \text{GlobalPooling}(\text{Concat}(\textbf{C}_2+\textbf{C}_4)) \\
         \textbf{w}_2, \textbf{w}_4 = \text{sigmoid}([f^1_{FC}(\textbf{v}), f^2_{FC}(\textbf{v})])\\
         \textbf{S} = \textbf{C}_2 \cdot \textbf{w}_2 + \textbf{C}_4 \cdot \textbf{w}_4,
    \end{array}
    \right.
\end{equation}
where $f^1_{FC}(\cdot)$ and $f^2_{FC}(\cdot)$ refer to two independent fully-connected layers. Finally, the final output $\textbf{J}$ can be obtained as follows:
\begin{equation}
    \textbf{J} =  \text{sigmoid}(f_{conv}(\textbf{S})) \cdot \textbf{I} + \textbf{I},
\end{equation}
where $f_{conv}$ is the function of one layer convolutional block.

\noindent \textbf{Non-local Attention.} \quad The detailed architecture of non-local attention is shown in Figure~\ref{fig:3}. Given the input $\textbf{I}$, we adopt three independent $1\times 1$ convolutional layers with trainable parameters as embedding functions. These three embedding functions are denoted as $g$, $\theta$, and $\phi$. They are used to generate the query (\textbf{Q}), key (\textbf{K}), and value (\textbf{V}). The embedding process does not change the size of the input, which can be defined as follows:
\begin{equation}
    \left\{
    \begin{array}{lr}
         \textbf{Q} = \theta(\textbf{I}) \\
         \textbf{K} = \phi(\textbf{I}) \\
         \textbf{V} = g(\textbf{I}).
    \end{array}
    \right.
\end{equation}
The distance matrix denoted by $\textbf{A}$ can be efficiently calculated as by dot product:
\begin{equation}
    \textbf{A} = \text{softmax}(\textbf{Q}\textbf{K}^{\top}).
\end{equation}
The output of non-local attention operation \textbf{W}, defined by Eq.(1), is calculated by the matrix multiplication between \textbf{A} and \textbf{V}. Finally, \textbf{W} is fed into a $1\times 1$ convolutional operation $o$ to generate the output $\textbf{J}$.

\subsection{Feature Filter}
The architecture of the feature filter is illustrated in Figure~\ref{fig:1}, which is the key component to achieve the selection operation. The feature filter contains three convolutional blocks, two LSTMs for capturing the correlations of the feature filter in different dynamic blocks, and a convolutional block followed by a sigmoid activation function. In addition, the selection operation is non-differentiable, thus we adopt Markov Decision Process (MDP) and reinforcement learning (RL) to train the feature filter. The state and action of this MDP and the developed difficulty-regulated reward for the feature filter training are clarified as follows:

\noindent \textbf{State and Action.} \quad In the $i$-th dynamic block, the state $s_i$ include the input feature $z_i$ and the hidden state of LSTM $h_i$. Given the state $s_i$, the feature filter produces a distribution of selection that can be expressed as $f_{AS} = \pi(\textbf{a}|s_i)$. In the training phase, the action is sampled from the probabilistic distribution, referred to as $\textbf{a}_i\sim \pi(\textbf{a}|s_i)$. While in the testing phase, the action is determined by the highest probability, i.e.,  $\textbf{a}_i=\text{argmax}_a \pi(\textbf{a}|s_i)$, in which the action $\textbf{a}_i$ is the two-channel mask that is described in Section 3.1.

\noindent \textbf{Difficulty-regulated Reward.} \quad In the RL framework, the feature filter is trained to maximize a cumulative reward. In this paper, for the attention selection, we develop a difficulty-regulated reward, which considers the network performance, complexity, and difficulty of the task. The reward at $i$-th dynamic block is expressed as:
\begin{equation}
r_i = \left\{
        \begin{array}{lr}
            -\gamma \times \textbf{a}_i,  & 1\leq i < N,\\
            -\gamma \times \textbf{a}_i + d \times (-\triangle L_2),  & i = N,
        \end{array}
\right.
\end{equation}
where $-\gamma$ is the reward penalty for selecting a path in one dynamic block. In addition, the performance gain in terms of L2 loss. The difficulty is denoted by $d$ is formulated as:
\begin{equation}
d = \left\{
    \begin{array}{lr}
        L_d/L_t,  & 0 \leq L_d < L_t, \\
        1, & L_d \geq L_t,
    \end{array}
\right.
\end{equation}
where $L_d$ is the Mean Square Error (MSE) loss function and $L_t$ is a threshold. As $L_d$ approaches zero, the difficult $d$ decreases, indicating the input noisy mixture is easy to enhance.

\begin{algorithm}[t]
\caption{REINFORCE algorithm}
\label{alg1}
    \begin{algorithmic}
    \STATE Get a batch of $K$ noisy-clean speech pairs, $\{y^{(1)}, s^{(1)}\}$, ..., $\{y^{(K)}, s^{(K)}\}$
    \STATE With policy $\pi$, derive ($s_1^{(k)}$, $a_{1}^{(k)}$, $s_1^{(k)}$, ..., $s_N^{(k)}$, $a_{N}^{(k)}$, $s_N^{(k)}$)
    \STATE Compute gradients using REINFORCE algorithm \cite{ref24}
    \STATE $\triangle \theta = \frac{1}{K} \sum\limits_{k=1}^K\sum\limits_{i=1}^N \nabla_{\theta}\log \pi(a_{i}^{(k)} | s_i^{(k)}; \theta)\sum\limits_{j=i}^N r_j^{(k)}$
    \STATE Update parameters $\theta \leftarrow \theta + \delta \triangle \theta$
    \end{algorithmic}
\end{algorithm}

\subsection{Implementation}
The Selector-Enhancer is composed of dual-attention based CNN and feature filter, and the training process of Selector-Enhancer consists of two stages. In the first stage, we train a dual-attention based CNN pre-trained model with a random policy mask. In the second stage, we train the feature filter and dual-attention based CNN together.

\noindent \textbf{Stage 1.} \quad We train the dual-attention based CNN pre-trained model with a randomly generated policy mask. In this study, we design a loss function by using MSE loss constrains the output to have reasonable quality. 

\noindent \textbf{Stage 2.} \quad We train the local and non-local based CNN and feature filter together. We train the feature filter using the REINFORCE algorithm \cite{ref24}, as illustrated in Algorithm 1. The parameters of the feature-filter and learning rate are denoted by $\theta$ and $\lambda$, respectively. 

\begin{figure}
\centerline{\includegraphics[width=\linewidth]{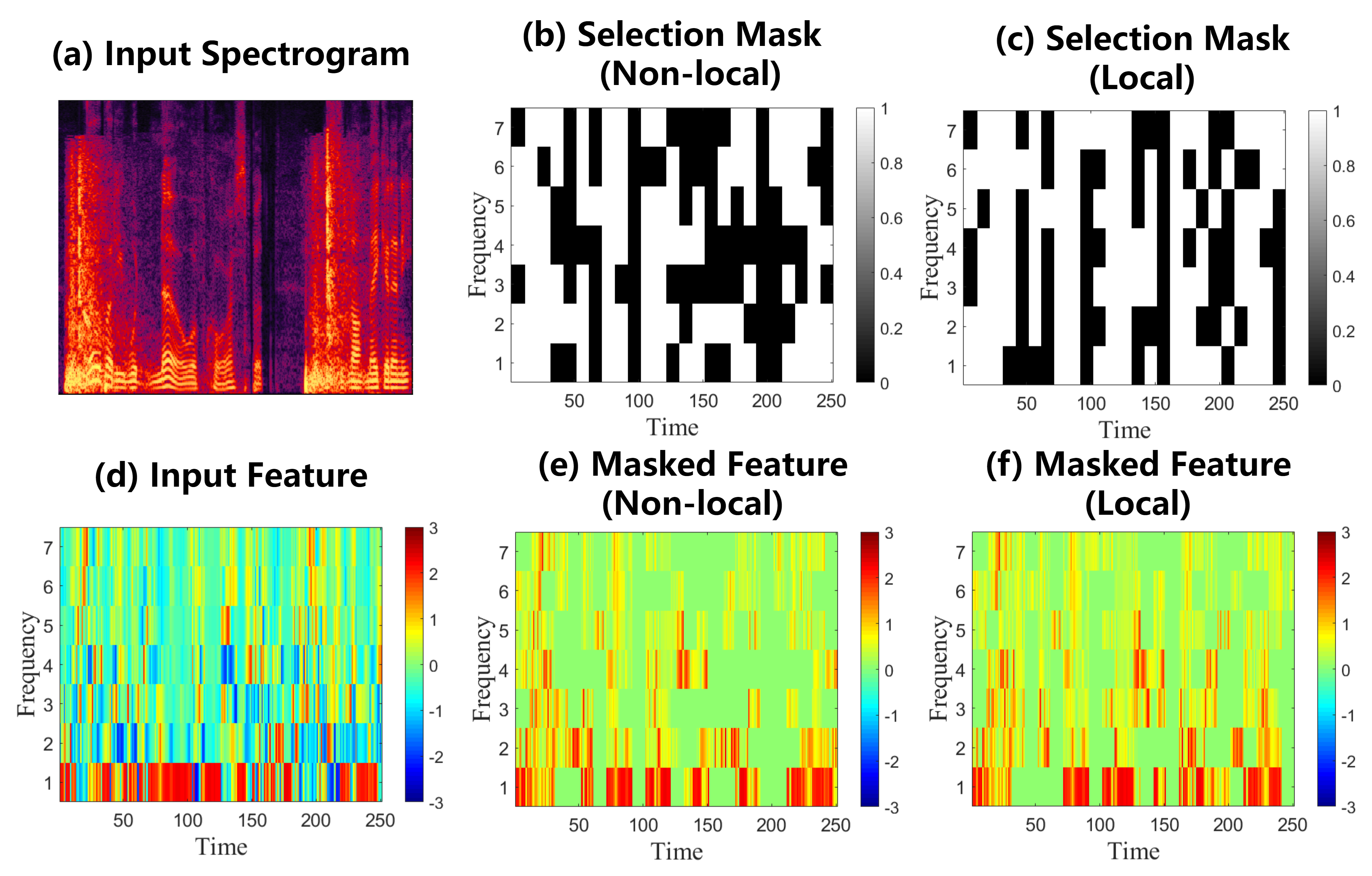}}
\caption{Visualization of the policy mask and masked feature by feature filter. Best viewed in color.}
\label{fig:4}
\end{figure}

\renewcommand{\arraystretch}{0.95}
\begin{table*}[t]
\centering
\caption{Ablation study of different components in Selector-Enhancer.}
\resizebox{1.0\textwidth}{!}{
\begin{tabular}{c|c|c|c|c|c|c|c|c|c|c}
\hline
Case Index          & 0 (default) & 1     & 2     & 3     & 4     & 5     & 6     & 7     & 8  & 9    \\ \hline
Local Attention     & $\checkmark$        & $\checkmark$  & $\times$ & $\checkmark$  & $\checkmark$  & $\times$ & $\checkmark$  & $\checkmark$  & $\checkmark$ & $\checkmark$ \\
Non-local Attention & $\checkmark$        & $\times$ & $\checkmark$  & $\checkmark$  & $\checkmark$  & $\times$ & $\checkmark$  & $\checkmark$  & $\checkmark$ & $\checkmark$ \\
Concatenation       & $\times$       & $\times$ & $\times$ & $\checkmark$  & $\times$ & $\times$ & $\times$ & $\times$ & $\times$ & $\times$ \\
Selective Network \cite{ref23}   & $\times$       & $\times$ & $\times$ & $\times$ & $\checkmark$  & $\times$ & $\times$ & $\times$ & $\times$ & $\times$ \\
Feature-Filter (Proposed)        & $\checkmark$        & $\times$ & $\times$ & $\times$ & $\times$ & $\times$ & $\checkmark$  & $\checkmark$  & $\checkmark$  & $\checkmark$ \\
Number of DB        & 4           & 4     & 4     & 4     & 4     & 4     & 3     & 2     & 1   & 5  \\ \hline
PESQ                & 3.09        & 2.63  & 2.66  & 2.82  & 2.99  & 2.36  & 3.02  & 2.94  & 2.87  & 3.11 \\
STOI ($\%$)               & 93.41       & 87.15 & 87.47 & 91.10 & 92.28 & 78.39 & 92.76 & 91.33 & 91.01 & 93.38 \\ \hline
Parameters          & 2.37M       & 1.98M & 2.27M & 2.38M & 2.55M & 2.15M & 2.24M & 2.03M & 1.81M & 2.57M \\ \hline
\end{tabular}}
\label{tab:1}
\end{table*}

\section{Experiments}
\subsection{Datasets}
To evaluate the proposed Selector-Enhancer, we use two types of datasets: VoiceBank + DEMAND and AVSpeech + AudioSet.

\noindent \textbf{VoiceBank + DEMAND:} \quad This is an open dataset proposed by \cite{ref29}. The pre-scaled subset of VoiceBank \cite{ref37} provided by is used to train the Selector-Enhancer. We randomly select 40 male speakers and 40 female speakers from the total of 84 speakers (42 male and 42 female), and each speaker pronounced around 4000 utterances. Therefore, the clean data contained 320,000 utterances in total (16,000 male utterances and 16,000 female utterances). Next, we mix 32,000 utterances with noise from DEMAND \cite{ref38} in $\{-5, -4, -3, -2, -1, 0, 5\}$ dB signal-to-noise ratio (SNR) levels. In addition, we set aside 200 clean utterances from the training set to create a validation set. Test data are generated by mixing 200 utterances selected from the remained 4 untrained speakers (50 each) with noise from DEMAND at three SNR levels $\{-5, 0, 5\}$ dB.

\noindent \textbf{AVSpeech+AudioSet:} \quad This is a large dataset proposed by \cite{ref39}. The clean dataset AVSpeech is collected from Youtube, containing 4700 hours of video segments with approximately 150,000 distinct speakers spanning a wide variety of people languages. The noisy speech is a mixture of the above clean speech segments with AudioSet \cite{ref40} that contains a total of more than 1.7 million 10-second segments of 526 kinds of noise. The noisy speech is synthesized by a weighted linear combination of speech segments and noise segments \cite{ref29}:
\begin{equation}
    \text{Mix}_i = \text{Speech}_j + 0.3 \times \text{Noise}_k,
\end{equation}
where $\text{Speech}_j$ and $\text{Noise}_k$ are 4-second segments randomly sampled from the speech and noise dataset. In our experiment, 10,000 segments were randomly sampled from the AVSpeech dataset for the training set and 500 segments for the validation dataset. Because of the wide energy distribution in both datasets, the created noisy speech dataset has a wide range of SNR.

For the test dataset, we randomly select 840 utterances from 84 speakers in WSJ0 SI-84 dataset \cite{ref41} and mingle with unknown types of non-stationary noises from NOISEX92 \cite{ref42} under three SNR levels $\{-5, 0, 5\}$ dB. The choice of the training set and test set is to let the Selector-Enhancer adapt to different speakers and different acoustic environments.

\subsection{Experimental Setup}

As for the number of dynamic blocks, we use $N = 4$ for a fair comparison with the baseline methods. The threshold of difficulty $L_t$ and the reward penalty $\gamma$ are, respectively, set to $6\times 10^{-2}$ and $8\times 10^{-2}$ based on the specific task. Specifically, when the noise corruption is severe, the threshold and reward penalty is required to be larger. In the first training stage, the coefficient of intermediate loss $\beta$ is set to 0.5. All the utterances are sampled at 16 kHz and features are extracted by using frames of length 512 with frame-shift of 256, and Hann windowing followed by STFT of size K = 512 with respective zero-padding. We train the model with a minibatch size of 16. We use Adam as an optimizer and implement our method on Pytorch. 

In our experiments, the performance is assessed by using standard speech enhancement metrics short-term objective intelligibility (STOI), and perceptual evaluation of speech quality (PESQ). STOI typically has a value range from 0 to 1 and can be interpreted as percent correct, and PESQ has a value range from -0.5 to 4.5.

\subsection{Ablation Study}
We show the ablation study in Table~\ref{tab:1} to investigate the effect of different components in Selector-Enhancer. Note that Case 0 denotes the proposed basic Selector-Enhancer with the default setting. In the ablation study, we  compare our method with several models: 1) \textbf{Case 1}: we remove all non-local attention mechanisms, 2) \textbf{Case 2}: we remove all local attention mechanisms, 3) \textbf{Case 3}: We remove the feature filters and adopt concatenation operation to fuse results of local and non-local operations, 4) \textbf{Case 4}: we remove feature filters and adopt selection network \cite{ref23} to fuse results of local and non-local operations, 5) \textbf{Case 5}: we remove all components of Selector-Enhancer. We show the detailed analysis of ablation below.

\noindent \textbf{Local Attention} \quad Compare with Case 3, Case 2 removes all local attention mechanisms. According to Table~\ref{tab:1}, we find that Case 3 improves PESQ by 0.16 and STOI by $3.63\%$ over Case 2. The obvious performance decrease in Case 2 indicates the positive effect of the local attention operation. In addition, we remove both local and non-local attention mechanisms in Case 5. Case 1 achieves 0.30 PESQ improvement and $9.08\%$ STOI improvement over Case 5. The performance comparison between Case 1 and Case 5 illustrates the necessity of the local operation for SE.

\noindent \textbf{Non-local Attention} \quad To verify the effect of non-local attention, we also provide two groups of comparison. Comparing Case 1 and Case 3, one can observe that local attention mechanisms produce 0.19 PESQ gains and $3.75\%$ STOI gains together with local attention mechanisms. What is more, From Case 1 to Case 5, the non-local attention mechanisms contribute 0.27 PESQ gains and $8.76\%$ STOI gains. Consequently, these two groups of comparison demonstrate the necessity of the non-local operation for SE.

\noindent \textbf{Attention Collaboration} \quad We declare that adopting local and non-local attention directly by deep learning networks is challenging since the speech contains dynamic and fast-changing features. To verify the necessity of attention collaborative operation, we make a comparison between Case 3 and Case 4. Note that Case 4 incorporates the local and non-local attention mechanisms by adopting selective network \cite{ref23}. From Case 3 to Case 4, the 0.17 PESQ gains and $1.18\%$ STOI gains by Case 4 over Case 3 fully demonstrate the superiority of the collaborative operation between local and non-local attention mechanisms.

\noindent \textbf{Feature-Filter} \quad To verify the necessity of our proposed feature filter, we compare Case 0 with Case 3 and Case 4 separately. The 0.27 PESQ gains and $2.31\%$ STOI gains by Case 0 over Case 3 and the 0.10 PESQ gains and $1.13\%$ STOI gains by Case 0 over Case 4 obviously indicate the effectiveness of our proposed feature filter for local and non-local attention collaboration. 

In order for a better understanding of the feature-filter, we visualize the masked feature maps processed by the feature filter for local and non-local attention mechanisms and processed feature maps produced by local and non-local attention operation in Figure~\ref{fig:4}. We show that the corrupted speech components are scaled and masked by feature-filter to separately feed into local and non-local attention operations. 

\noindent \textbf{Number of Dynamic Blocks} \quad To study the number of dynamic blocks, we make a comparison among Cases 0, 6, 7, 8, and 9, which clearly shows that the performance increases with the number of dynamic blocks. Therefore, for a good trade-off between speed and accuracy, we use 4 dynamic blocks as our final proposed model to do comparisons to other state-of-the-art works.

\renewcommand{\arraystretch}{1.05}
\begin{table*}[t]
\centering
\caption{Evaluation and comparison of different models on VoiceBank + DEMAND.}
\resizebox{1.00\textwidth}{!}{
\begin{tabular}{cccccccccc}
\hline
SNR               & \multicolumn{2}{c}{-5 dB}      & \multicolumn{2}{c}{0 dB}       & \multicolumn{2}{c}{5 dB}       & \multirow{2}{*}{FLOPs (G)} & \multirow{2}{*}{Param. (M)} & \multirow{2}{*}{RTF} \\ \cline{1-7}
Metrics           & STOI (\%)      & PESQ          & STOI (\%)      & PESQ          & STOI (\%)      & PESQ          &                            &                             &                      \\ \hline
Unprocessed       & 56.32          & 1.30          & 70.95          & 1.57          & 81.33          & 1.83          & -                          & -                           & -                    \\ \hline
CRN (2019)        & 73.36          & 1.81          & 82.27          & 2.45          & 89.27          & 2.90          & 1.16                       & 2.36                        & 0.52                 \\
CNN-NL (2020)     & 77.98          & 1.90          & 85.63          & 2.51          & 90.11          & 3.05          & 5.94                       & 5.04                        & 0.86                 \\
T-GSA (2020)      & 80.88          & 2.14          & 88.47          & 2.66          & 91.37          & 3.03          & 6.23                       & 5.63                        & 0.95                 \\
SA-TCN (2021)     & 82.47          & 2.26          & 89.89          & 2.84          & 92.73          & 3.25          & 8.85                       & 5.81                        & 0.91                 \\ \hline
Selector-Enhancer & \textbf{83.38} & \textbf{2.36} & \textbf{91.24} & \textbf{3.03} & \textbf{94.36} & \textbf{3.37} & 2.19                       & 2.37                        & 0.63                 \\ \hline
\end{tabular}}
\label{tab:2}
\end{table*}

\subsection{System Comparison}
We carry out the system comparison on both datasets mentioned in section 4.1.

\noindent \textbf{VoiceBank+DEMAND:} \quad We train the proposed Selector-Enhancer on the synthetic and commonly used dataset VoiceBank+DEMAND. The results are provided in Table~\ref{tab:2}. We compare our model with 4 recent baselines, CRN \cite{ref28}, CNN-NL \cite{ref10}, T-GSA \cite{ref15}, and SA-TCN \cite{ref31}. Firstly, the non-local attention based SE models like CNN-NL, T-GSA, and SA-TCN have a very large gain over the simple CNN-based SE model (CRN) in terms of STOI and PESQ. This indicates the advantage of non-local attention operations in capturing long-term dependencies. Also, Selector-Enhancer yields the best results in all conditions.

Realizing that the number of trainable parameters can not completely reflect the complexity of a model, especially for deep non-local methods. Thus we employ the floating-point operations (FLOPs), the number of trainable parameters, and real-time factor (RTF) to evaluate the complexity of different methods. The complexity analysis of several representative methods is reported in Table~\ref{tab:2}. The proposed Selector-Enhancer has an attractive complexity and low computational costs and achieves quite promising performance.

\renewcommand{\arraystretch}{1.05}
\begin{table}[t]
\centering
\caption{Performance evaluation scores on AVSpeech + AudioSet}
\resizebox{0.48\textwidth}{!}{
\begin{tabular}{ccccccc}
\hline
Test SNR   & \multicolumn{2}{c}{-5 dB}      & \multicolumn{2}{c}{0 dB}       & \multicolumn{2}{c}{5 dB}       \\ \hline
Metrics    & STOI      & PESQ          & STOI      & PESQ          & STOI      & PESQ          \\ \hline
Mixture    & 57.99          & 1.18          & 57.99          & 1.18          & 90.03          & 2.06          \\ \hline
U-Net & 72.92          & 1.76          & 87.01          & 2.06          & 91.85          & 2.37          \\
GRN       & 74.96          & 1.81          & 88.62          & 2.19          & 91.72          & 2.44          \\
PHASEN    & 74.77          & 2.47          & 89.89          & 2.93          & 93.70          & 3.09          \\
TFT-Net   & 76.94          & 2.47          & 90.33          & 2.93          & 93.70          & 3.11          \\
Selector-Enhancer  & \textbf{78.33} & \textbf{2.52} & \textbf{91.19} & \textbf{3.06} & \textbf{94.27} & \textbf{3.15} \\ \hline
\end{tabular}}
\label{tab:3}
\end{table}

\noindent \textbf{AVSpeech+AudioSet:} \quad On this large dataset, we compare the Selector-Enhancer with several state-of-the-art methods,i.e., U-Net \cite{ref3}, GRN \cite{ref7}, PHASEN \cite{ref29}, and TFT-Net \cite{ref30}. The results in Table~\ref{tab:3} demonstrate that the Selector-Enhancer outperforms these state-of-the-art methods and also indicate that the Selector-Enhancer can be generalized to various speakers and various kinds of noisy environments. 

\begin{figure}[t]
\centering
\subfigure[STOI score ($\%$)]{
\label{Fig7.sub.1}
\includegraphics[width=0.223\textwidth]{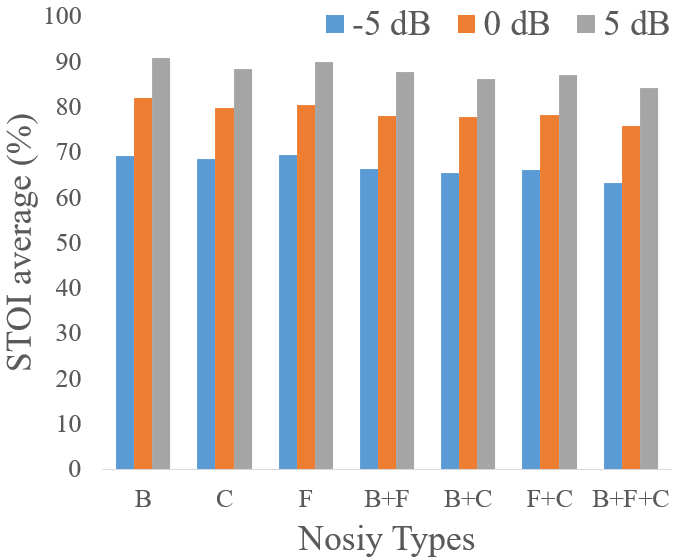}}
\subfigure[PESQ score]{
\label{Fig7.sub.2}
\includegraphics[width=0.223\textwidth]{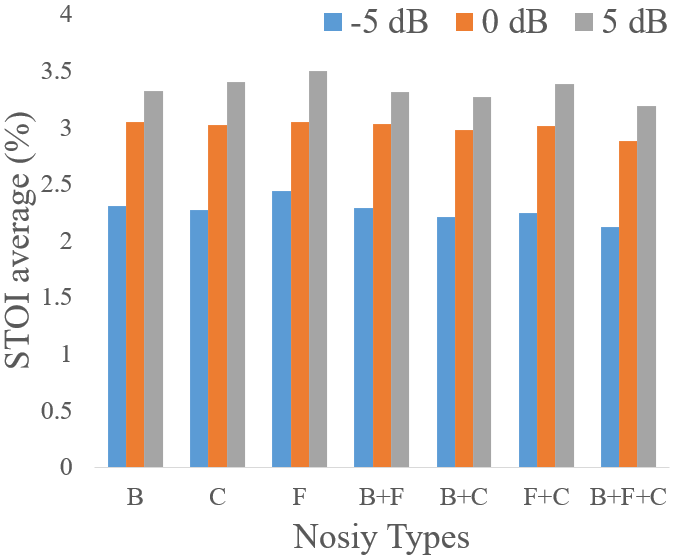}}
\caption{Comparison in STOI and PESQ average for cases with multiple noise types, in which B, F, and C denote babble, factory, and cafeteria noisy types, respectively.}
\label{fig:7}
\end{figure}

To demonstrate the effectiveness and flexibility of the Selector-Enhancer, we further explore and apply the Selector-Enhancer to circumstances when multiple noise types are included in a noisy background. We randomly select 150 clean utterances from the test set and 3 challenging noises, babble (B), factory (F), and cafeteria (C) from NOISEX92, which are explained in Section 4.1, to produce the multiple types of noises with clean utterances at -5 dB, 0 dB, and 5 dB SNR levels. Figure~\ref{fig:7} shows the STOI and PESQ results over three mentioned SNR levels on multiple noise types, where B, F, and C represent babble, factory, and cafeteria noise types, respectively. According to the Table, we observe that the results using 1, 2, and 3 noise types in the noisy background are comparable, although the STOI and PESQ scores are decreasing when the number of noise types is increase.

\section{Conclusion}
In this paper, we propose Selector-Enhancer that enables attention selection for each speech feature region. Specifically, Selector-Enhancer contains a dual-attention based CNN and a feature filter. The dual-attention based CNN offers options for local and non-local attention mechanisms to address different types of noisy mixtures while the feature-filter selects the optimal regions of speech spectra to feed into the specific attention mechanisms. Comprehensive ablation studies have been carried out, justifying almost every design choice we have made in Selector-Enhancer. Experimental results show that the proposed method is capable of superior performance in SE by outperforming the state-of-the-art methods with fewer computational costs.

\bibliography{aaai23}

\end{document}